\documentclass[twocolumn,amsmath,pre,floatfix,superscriptaddress]{revtex4}
\usepackage{graphicx}% Include figure files
\usepackage{bm}% bold math
\usepackage{amsmath}% More mathematical features
\usepackage{amssymb}     
\usepackage{epsfig}
\usepackage{pstricks,pst-grad}% PsTricks
\usepackage{bbm}

\begin{document}

\title{Effect of the reservoir size on gas adsorption in inhomogeneous porous media}

\author{E.~Kierlik}
%\email{mlr@lptmc.jussieu.fr}
%
\affiliation{Laboratoire de Physique Th\'eorique de la Mati\`ere Condens\'ee, Universit\'e Pierre et Marie Curie\\ 4 place Jussieu, 75252 Paris Cedex 05, France}

\author{J.~Puibasset}
%
%\affiliation{Laboratoire de M\'ecanique des Solides, Ecole Polytechnique, 91128 Palaiseau, France}
\affiliation{Centre de Recherche sur la Mati\`ere Divis\'ee, Universit\'e d'Orl\'eans\\ 1b rue de la F\'erollerie, 45071 Orl\'eans Cedex 02, France}

\author{G.~Tarjus}
\affiliation{Laboratoire de Physique Th\'eorique de la Mati\`ere Condens\'ee, Universit\'e Pierre et Marie Curie\\ 4 place Jussieu, 75252 Paris Cedex 05, France}
%
%\email{tarjus@lptmc.jussieu.fr}

\date{\today}

\begin{abstract}
We study the influence of the relative size of the reservoir on the adsorption isotherms of a fluid in disordered or inhomogeneous mesoporous solids. We consider both an atomistic model of a fluid in a simple, yet structured pore, whose adsorption isotherms are computed by molecular simulation, and a coarse-grained model for adsorption in a disordered mesoporous material, studied by a density functional approach in a local mean-field approximation. In both cases, the fluid inside the porous solid exchanges matter with a reservoir of gas that is at the same temperature and chemical potential and whose relative size can be varied, and the control parameter is the total number of molecules present in the porous sample and in the reservoir. Varying the relative sizes of the reservoir and the sample may change the shape of the hysteretic isotherms, leading to a ``reentrant'' behavior compared to the grand-canonical isotherm when the latter displays a jump in density. We relate these phenomena to the organization of the metastable states that are accessible for the adsorbed fluid at a given chemical potential or density.
\end{abstract}

\maketitle

\def\be{\begin{equation}}
\def\ee{\end{equation}}
\def\bea{\begin{align}}
\def\eea{\end{align}}

\section{Introduction}
Under confinement in disordered mesoporous materials, the characteristic time scale for relaxation of a fluid can become
extremely long. As a result, and although not always appreciated, equilibrium is often not attained and, accordingly, \textit{bona fide}
thermodynamic transitions such as the liquid-gas transition in one-component fluids or macroscopic phase separation in mixtures, are
unobservable. ``Capillary condensation'' in disordered solids is an out-of-equilibrium phenomenon, as illustrated by the irreversibility
and hysteresis effects found in experiments. One typically observes a hysteresis loop that describes the isothermal evolution of
the amount of fluid adsorbed in the porous solid as a function of the applied pressure, with branches that differ on adsorption (filling) and on desorption (draining). This hysteresis loop appears rate-independent, and its size and shape vary with temperature as well as with the characteristics of the solid (\textit{e.g.} its porosity) or those of the solid-fluid interaction potential. Such a phenomenon is related to the existence of a large number of ``metastable states'' in which the system can be trapped on the experimental time scale; evolution from one metastable state to another then only occurs as a result of the action of the applied pressure (or equivalently, chemical potential) and it proceeds through a sequence of irreversible cooperative condensation (or evaporation) events, generically denoted as ``avalanches''.\cite{DKRT2006} The fact that the location and the shape of the hysteresis loop are reproducible in experiments indicates that the observation time is smaller than the time to reach the global equilibrium state, but is larger than local equilibration processes by which the system settles in one metastable state. As a result, the behavior of a fluid during filling or draining in disordered mesoporous materials can be rationalized by envisaging the evolution of the system in a free-energy landscape characterized by many local minima, \textit{i.e.} metastable states.\cite{KMRST2001,KMRT2002,WM2003}

The above picture of gas adsorption in disordered porous media brings in a strong analogy with the out-of-equilibrium response of systems driven by an external force in the presence of impurities or other types of quenched disorder. This is for example the case of
magnetization cycles in ferromagnetic materials when a magnetic field is ramped up and down and of hysteretic martensitic transformations in alloys;\cite{B2006} in both examples, avalanches can be detected through some ``crackling noise'',\cite{SDP2004} magnetic Barkhausen noise in the former, acoustic emission in the latter. In such driven disordered systems, one expects the occurrence of out-of-equilibrium phase transitions as one changes, on top of the driving force, some external parameters such as the temperature or the characteristics of the intrinsic disorder (\textit{e.g.} the porosity in a porous solid).\cite{SDKKRS1993} The branches of the hysteresis loop, in particular the adsorption and the desorption isotherms, may then display jumps (discontinuities): indications for such behavior are for instance seen in the adsorption of helium in very light aerogels.\cite{TYC1999,HDB2005,BLCGDPW2008}

Such discontinuities and out-of-equilibrium phase transitions, however, are theoretically predicted on the basis of a grand-canonical set-up (for gas adsorption) in which the gas reservoir is infinite. In real experiments, the reservoir has a finite size which may not
always be large enough for considering that the fluid inside the porous material is in a grand-canonical ensemble with fixed chemical
potential. What should be expected in such situations ? The evolution of the system among metastable states may depend on the specific experimental set-up. As far as we know, there has been no systematic experimental study of the effect of changing the size of the gas reservoir (relative to that of the porous medium) and no estimate of the condition under which a grand-canonical situation is approximately reached. Answering these questions is the primary goal of the present paper.

The question of the dependence of the hysteretic response on the chosen control parameter has been a little more studied in other driven
disordered systems. Two extreme cases have been considered: the response of an extensive quantity to a change in the (conjugate) ``force'' and the response of an intensive quantity (a force) to a change in the conjugate extensive quantity; the former
is akin to the grand-canonical situation for a fluid in which the chemical potential (or actually, the pressure in the gas reservoir) is
controlled and the latter to a canonical situation in which the number of adsorbed fluid molecules is controlled. Examples of such studies are found in the context of martensitic transformations in which either stress or strain is controlled\cite{BRIMPV2007} and in that of Barkhausen noise in which either magnetic field or, via some feed-back mechanism, magnetization is controlled.\cite{B2006} In all cases, the loop obtained with the extensive variable as control parameter appears as reentrant when compared to that obtained with the force as control parameter. (For a theoretical study, see Ref. \cite{IRV2006})

One anticipates that the influence of the relative size of the reservoir\cite{fn1} on the adsorption isotherms of a fluid in a disordered porous material is intimately connected to the organization of the metastable states in the adsorbed-density/chemical-potential plane. Indeed, when the isotherms are smooth both on adsorption and on  desorption, one may experimentally probe metastable states located inside the main hysteresis loop by studying ascending and descending ``scanning curves'', which involve partial filling or draining.\cite{B2006} These curves lead to a variety of hysteretic subloops that provide direct evidence  for the presence of metastable states inside the main loop.\cite{BE1989,BEN1989,LFH1993} Actually, metastable states are expected everywhere inside the latter. The situation is quite different, however, when the adsorption or the desorption branch displays a jump (in a grand-canonical setting with a very large reservoir). A discontinuity prevents the realization of scanning curves in some portion of the isotherm. As a result, part of the interior of the main hysteresis loop is now inaccessible.

In the present work we address the above questions, namely the effect of the relative size of the gas reservoir on the adsorption isotherms of a fluid in a disordered or inhomogeous porous solid and the connection to the distribution of metastable states inside the hysteresis loop. We show that even when the grand-canonical isotherms display discontinuous jumps, there are branches of metastable states inside the main hysteresis loop and that many of these states can be reached by varying the relative size of
the reservoir, the smaller the reservoir the larger the extent to which the branches of metastable states are probed. In particular, the
extent is maximal when the reservoir is so small that the fluid inside the porous solid behaves as in a canonical situation of fixed number of adsorbed molecules. When compared to the loop obtained for an infinite reservoir (grand-canonical situation) and plotted as the amount adsorbed versus the chemical potential, the hysteresis loop for a finite reservoir size then appears as reentrant. On the contrary, when the isotherms are smooth (continuous), there is no influence of the size of the reservoir.

The rest of the paper is organized as follows:

In section II, we introduce the two models that have been investigated and give some details on the methods used to compute the adsorption isotherms. We have considered: (i) an atomistic model of a fluid in simple, yet structured pore, whose adsorption isotherms are computed by molecular simulation, and (ii) a coarse-grained model for adsorption in a disordered mesoporous material, studied by a density functional approach in a local mean-field approximation. In both cases, the fluid inside the porous solid exchanges matter with a reservoir of gas that is at the same temperature and chemical potential and whose relative size can be varied. The overall system composed of the sample plus the reservoir is taken in the canonical ensemble, with the total number of molecules as control parameter.

In section III, we present the results for the atomistic model. The simplicity of the system allows one to get a clear interpretation of
the physical nature of the branches of metastable states, whose number is limited. The existence of metastable states is a direct consequence of the intrinsic inhomogeneity of the pore space. As the relative size of the gas reservoir is varied, we find that the exploration of the branches changes, the isotherm displaying larger jumps in the adsorbed density as the size increases.

Section IV is devoted to the coarse-grained model of a disordered porous material. The number of metastable states is now enormous,
since it is likely to increase exponentially with the size of the sample. We find a drastic change of behavior between the ``strong-disorder'' regime for which the isotherm is continuous and the ``weak-disorder'' one for which the isotherm, here the adsorption isotherm on which we focus our study, displays a jump corresponding to a macroscopic avalanche. In the former regime, the isotherm does not display any dependence on the relative size of the reservoir whereas in the latter, it is reentrant for finite reservoir sizes, the more so as one decreases the reservoir size, and it shows increasing jumps in adsorbed density with increasing reservoir size.

Finally, we summarize our main results and give our conclusions in section V. In particular, we discuss the relevance of our study to experimental situations and we stress the important role of the intrinsic inhomogeneity induced by the solid matrix.

\section{Models and methods}

\subsection{the set-up: sample plus reservoir}
The principles of the method are the following. One considers a starting situation where $N$ particles are placed inside two cells which are in thermal equilibrium with an infinite heat bath at temperature $T$. One of the cells represents a porous solid sample of volume $V_P$ and the other a reservoir of volume $V_R$. The total volume is then equal to $V= V_R + V_P$. We allow mass exchange between the cells whose volumes are kept fixed so that at (exchange) equilibrium the chemical potential $\mu_R$ of particles inside the reservoir is equal to the chemical potential $\mu_P$ of particles inside the porous solid: $\mu=\mu_R=\mu_P$. We measure the average amount of particles $N_P$ present in the porous solid. Varying the total number of particles $N$ in small steps, one changes both the chemical potential of the cells, $\mu(N)$, and the number of adsorbed particles, $ N_P(N)$. We then monitor the adsorption isotherm $\rho= N_P / V_P$ (for simplicity, we omit the subscript $P$ in the fluid density adsorbed in the porous sample) as a function of $\mu$. Introducing the ratio $\alpha= V_P /V$ and the average reservoir density $\rho_R = N_R / V_R$, where $N_R=N-N_P$, one has to satisfy for each $\mu$ the constraint $\rho_T=(N/V)=\alpha \rho+(1-\alpha)\rho_R$  with the condition $\mu =\mu_R(\rho_R,T)=\mu_P(\rho,T)$. Taking the limit $\alpha \rightarrow 0$  corresponds to the grand-canonical ensemble for the adsorbed fluid (with a controlled chemical potential) whereas taking the limit to 1 corresponds to the canonical ensemble for the adsorbed fluid (with a controlled number of adsorbed particles). Between these two extremes, one has a mixed ensemble, with a reservoir of variable size compared to the sample, somewhat reminiscent of the mesoscopic canonical ensemble considered by Neimark and coworkers.\cite{N2000} More details about the algorithms used in our two models are given below.

\subsection{Atomistic model of a fluid in a single structured pore}
An atomistic model of confined fluid was considered because it allows a realistic description of the various (metastable) states the system can adopt and, possibly, of the transition mechanisms at a molecular scale.
The chosen system is a simple atomic fluid (Lennard-Jones like) confined in a mesoporous substrate. (The model more specifically corresponds to Argon adsorbed in nanoporous solid carbon dioxide.)\cite{PWG1986,PGH1988} A fully realistic model of the substrate should take into account the surface roughness, pore morphology (pore size distribution), and interconnections between pores. The number of metastable states would however be too large for a systematic study. The model was therefore designed to exhibit only a few metastable states.

\begin{figure}[ht]
\epsfig{file= 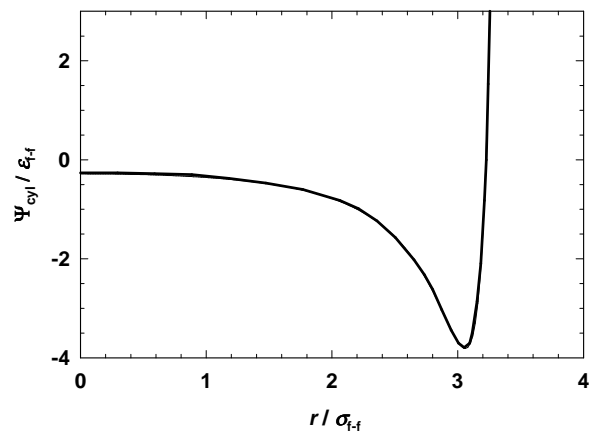, width=8cm,clip=}
\caption{External potential $\Psi^*_{cyl}(r)$ for the perfectly cylindrical pore as a function of radial distance $r$. }
\label{Figpotr}
\end{figure}

\begin{figure}[ht]
\epsfig{file= 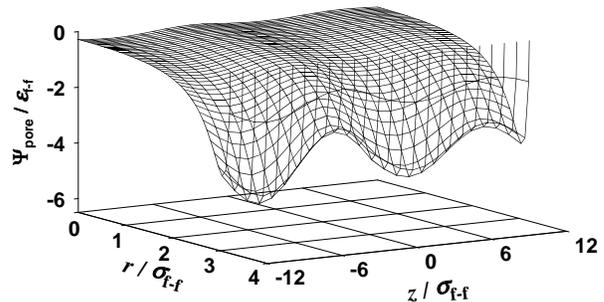, width=8cm,clip=}
\caption{External potential $\Psi^*_{pore}(r,z)$ for the heterogenous pore as a function of coordinates $r$ and $z$.}
\label{Figpotrz}
\end{figure}

The main potential sources for generating metastable states are nanometer-scale heterogeneities, due to pore-size distribution, and variations in surface chemistry. In both cases, the prominent effect on the adsorbed fluid comes from the modulation of the effective fluid/wall interaction.\cite{P2005a,P2005b} We then chose to investigate a cylindrical pore containing few domains of variable fluid/wall interaction.\cite{P2006} 
The diameter of the nanopore is 8 $\sigma_{ff}$  ($\sigma_{ff}$ is the fluid-fluid Lennard-Jones diameter). In this way, the smooth-wall approximation can be applied: the external potential seen by a fluid particle in the pore is calculated by integrating the fluid-wall ($6-12$) Lennard-Jones potential ($\epsilon_{sf}$ = 1.277 $\epsilon_{ff}$, $\sigma_{sf}$ = 1.094 $\sigma_{ff}$) over a uniform distribution of substrate interacting sites of density 0.8265 $\sigma_{ff}^{3}$. The calculated reduced external potential $\Psi^*_{cyl}(r)=\Psi_{cyl}(r)/ \epsilon_{ff}$ in this perfectly cylindrical pore is given in Fig. \ref{Figpotr}. The heterogeneity is introduced by modulating this external potential along the axial direction z:
\be 
\label{Eq01}
\Psi^*_{pore}(r,z)=\left[ 1+a(z) \sin \left(\frac{4\pi z}{L}\right)\right] \Psi^*_{cyl}(r)
\ee
where $a$ = 0.3 for $z <0$ and $a$ = 0.2 for $z >0$, and $L= 24\sigma_{ff}$  is the simulation box length. This external potential exhibits four domains of spatial extension of a few molecular diameters along the axial direction (see Fig. \ref{Figpotrz}). Periodic boundary conditions are applied along the axial direction $z$. The interactions are truncated at half the simulation box size (minimal image convention).

The fluid adsorption properties are calculated by Monte-Carlo simulations. Thermalization of the adsorbed fluid is performed by particle displacement trials. Chemical equilibration between the adsorbed fluid and the reservoir is done by particle exchange trials. The acceptance probabilities are given by the Metropolis algorithm. The fluid in the reservoir is assumed to be ideal, which is a good approximation since the simulations are performed well below the critical point. As a consequence, the gas in the reservoir does not need to be treated explicitly, which speeds up the calculations. In the limit of infinite reservoir size, one recovers the usual Grand Canonical Monte Carlo algorithm (GCMC). The adsorption/desorption curves are determined as follows. The initial configuration is the empty pore. Few molecules are then added in one step to the system (pore+reservoir). In order to mimic the experimental situation, the molecules are actually added to the reservoir cell, which results in an increase in $\mu_{R}$, and, accordingly, to a lack of equilibrium between the pore and its reservoir. The whole system is then relaxed for a while. Molecules exchange between the cells, which produces a small flow from the reservoir to the pore in the first steps. This flow comes to zero when stationarity is reached ($\mu_{R} = \mu_{P}$). We emphasize that this matter flow does not describe correctly the mass transport, since the Monte Carlo algorithm allows transfer of matter everywhere in the porous substrate. $10^6$ Monte Carlo trials per particle are then performed to acquire statistics for computing averaged quantities. This gives the first point of the isotherm. The subsequent points are obtained according to the same procedure, \textit{i.e.} a small increase in the number of molecules in the reservoir, followed by a relaxation run, and, finally, by a long run for acquisition. After complete adsorption has been achieved, the desorption isotherm can be calculated by decreasing the amount of particles in the reservoir step by step, as in a real experiment. This algorithm allows us to obtain the scanning curves embedded within the main hysteresis loop. Note that for the case of an infinite reservoir, it is the chemical potential which is increased stepwise instead of the amount of particles in the reservoir, which is then infinite.

\subsection{Coarse-grained model for adsorption in a disordered porous material}
As discussed in previous papers,\cite{KMRST2001,SM2002,KMRT2002} our approach to fluid adsorption in disordered mesoporous materials is based on a coarse-grained lattice-gas description which incorporates the essential physical ingredients of the solid-fluid system. We consider a three-dimensional BCC-lattice of linear size $L$ in which each of the $N_{SP}=2L^3$ sites may be occupied by a fluid or by a solid particle ($L$ is measured in units of the lattice spacing $a$ and we set $a\equiv 1$; for a lattice, the volume $V_P$ is then simply equal to the number of sites $N_{SP}$ times the volume of the elementary cell $a^3\equiv1$, so that we shall speak of volume and number of sites indistinctly). Multiple occupancy of a site is forbidden and only nearest-neighbor interactions are taken into account. The fluid particles can equilibrate, as explained below, with a finite-size reservoir at fixed temperature whereas the solid particles are ``quenched'' and distributed according to a specific choice of the porous material structure. For the sake of simplicity, we study a random matrix with a porosity (i.e. fraction of sites without solid particles) $\phi=75$\%. The relevant correlation length of the solid is around one lattice spacing so that even the smallest systems studied (with a linear size $L=25$) can correctly describe the collective effects occurring inside the matrix on long length scale (such as a sharp condensation event in the whole pore space).

The starting point of our theoretical analysis is the following expression of the free-energy functional in the mean-field approximation:
\begin{align}
\label{Eq02} 
&F_P[\{ \rho_i \}]=k_BT \sum_i [\rho_i \ln \rho_i +(\eta_i -\rho_i) \ln (\eta_i -\rho_i)  ] \nonumber \\
&-w_{ff} \sum_{<ij>} \rho_i \rho_j-w_{sf} \sum_{<ij>} [\rho_i (1-\eta_j)+\rho_j (1-\eta_i)]
\end{align}
where $\rho_i  $ is the thermally averaged fluid density at site $i$ and $\eta_i =0,1$ is a quenched variable describing the occupation of the lattice by solid particles ($\eta_i =0$ if the site $i$ is occupied by the solid and $\eta_i =1$ otherwise); $w_{ff}$ and $w_{sf}$ denote the fluid-fluid and fluid-solid attractive interactions, respectively, and the double summations run over all distinct pairs of nearest-neighbor sites.

We first start with the grand canonical situation ($\alpha=0$) where fluid particles can equilibrate with an infinite reservoir that fixes the chemical potential $\mu$. Minimizing the grand-potential functional  $\Omega_P[\{ \rho_i \}]= F_P[\{ \rho_i \}]-\mu \sum_i \rho_i$ with respect to $\{ \rho_i \}$ at fixed $T$ and $\mu $ for a given realization of the solid yields a set of  coupled equations,
\be
\label{Eq03} 
\rho_i=\frac{\eta_i}{1+\exp [ -\beta (\mu +w_{ff}\sum_{j/i} \rho_j+w_{sf}\sum_{j/i} (1-\eta_j))]},
\ee
where the sums run over the $c=8$ nearest neighbors of site $i$.  By using a simple iterative method to solve these equations, one finds solutions that are only minima of the grand potential surface, \textit{i.e.} metastable states.
For a given realization of the solid, the adsorption isotherm is obtained by increasing the chemical potential in small steps $\delta \mu$. At each subsequent $\mu $, the converged solution at $\mu-\delta \mu$  is used to start the iterations. 

What happens in the mixed situation ($\alpha \neq 0$)? The fluid particles can equilibrate between the solid sample and a finite-reservoir so that their total number $N$ is fixed. The isotherm is then obtained by increasing $N$ in small steps $\delta N$. In this case, the system tries to minimize its total Helmholtz free-energy $F_T[\{ \rho_i \},\rho_R,T]=F_P[\{ \rho_i \},T]+F_R[\rho_R,T]$, where $F_R$ is the Helmholtz free-energy of the gas reservoir  ($F_R[\rho_R,T]=V_R\{k_BT [\rho_R \ln \rho_R +(1 -\rho_R) \ln (1 -\rho_R)  ] -w_{ff}c\rho_R^2/2\}$  with $V_R$ the volume of the reservoir, \textit{i.e.}, again up to $a^3\equiv1$, the number of sites of the reservoir, while satisfying the global constraint $N=N_P+N_R=\sum_i \rho_i+V_R \rho_R$. This can be solved in a natural way by the method of Lagrange multipliers. We consider the function
\be
\label{Eq04} 
\bar{\Omega}_T[\{ \rho_i \},\rho_R,\lambda ,T]=F_T[\{ \rho_i \},\rho_R,T]+\lambda \{ N-\sum_i \rho_i-V_R \rho_R\},
\ee
where $\lambda $ is a Lagrange multiplier that has the meaning of the chemical potential coupled to the densities. Minimizing $F_T$ with the constraint on densities amounts to solve simultaneously the coupled equations $\frac{\partial \Omega_T}{\partial \rho_i}=0$, $\frac{\partial \Omega_T}{\partial \rho_R}=0$ and  $\frac{\partial \Omega_T}{\partial \lambda}=0$
or equivalently
\begin{align}
\label{Eq05} 
& k_B T \ln [\frac{\rho_i}{\eta_i-\rho_i}]-\lambda-w_{ff}\sum_{j/i} \rho_j - w_{sf}\sum_{j/i} (1-\eta_j)= 0, \nonumber \\
& k_BT \ln [\frac{\rho_R}{1-\rho_R}]-\lambda-w_{ff}\rho_R=0, \nonumber \\
& N-\sum_i \rho_i-V_R \rho_R=0, \quad 1 \leq i \leq N_{SP}.
\end{align}
One has then to define an iterative scheme that specifies how the system goes from one converged solution to another as the total number of particles is slowly changed. The details are given in the appendix.

We are searching for configurations for the local densities ${\rho_i}$ that are local extrema of the grand-potential functional $\Omega_P$ for the special value of the chemical potential $\mu=\lambda$. However, it is not fully guaranteed that the above algorithm necessarily converges to a local minimum, nor even to an extremum, since the constraint could in principle stabilize an unstable state. Therefore, we regularly ascertained that configurations obtained in the mixed ensemble are stable in the grand-canonical ensemble, \textit{i.e.} that they indeed are metastable states, by starting grand-canonical calculations with the converged solution at chemical potential $\lambda $. In addition, it must be emphasized that the global constraint is not satisfied until convergence is reached, in the spirit of the Lagrange method. It is therefore doubtful that one can attribute any physical meaning to the intermediate stages of the iterative process. Recall also that one does not take into account mass transport: local densities can change everywhere and instantaneously inside the porous sample.

Moreover, we focus on the adsorption (filling) process. As shown in previous papers,\cite{KMRT2002,RKT2003} fluid desorption may crucially depend on the presence of an external surface for the porous solid: the system then includes a real interface between the adsorbed fluid and the external vapor, and, during desorption, the vapor domain may penetrate and drain the solid from the outer surface (the so-called percolation and depinning transitions discussed in Ref. \cite{RKT2003}. These mechanisms are no longer bulk phenomenon and their, \textit{a priori} more subtle, analysis will be carried out elsewhere. We have therefore used periodic boundary conditions for the sample and the reservoir (separately). The ratio $y= w_{sf}/w_{ff}$ has been fixed to $0.8$ so that the adsorption isotherms exhibit the interesting range of phenomena as the temperature changes (see Ref.\cite{KMRT2002}).

\section{Results on reservoir-size dependence for the atomistic model}

\subsection{Grand-canonical isotherms}
\begin{figure}[ht]
\epsfig{file= 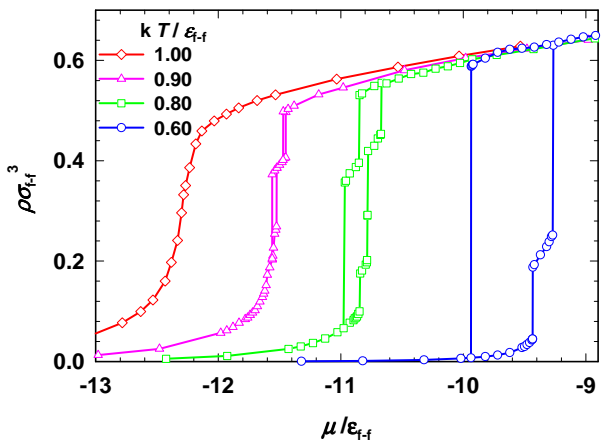, width=8cm,clip=}
\caption{Adsorption/desorption isotherms obtained by GCMC simulation for various temperatures (symbols). Lines are guides to the eye. }
\label{Figisothmain}
\end{figure}

We first focus on the adsorption/desorption isotherms obtained in the limit of infinite reservoir size ($\alpha =0$), \textit{i.e.} the GCMC data. The results are given in Fig. \ref{Figisothmain} for various temperatures. As can be seen, the highest temperature isotherm (reduced temperature $T^* =k_BT/\epsilon_{ff}= 1.0$) is perfectly reversible, \textit{i.e.} the adsorption and desorption curves are superimposed. On the other hand, for lower temperatures, the adsorption and desorption curves differ and exhibit hysteresis with vertical steps. We emphasize that the vertical lines are guides to the eye and do not correspond to GCMC data. The isotherm jumps in one step, and cannot be stabilized in between.

\begin{figure}[ht]
\epsfig{file= 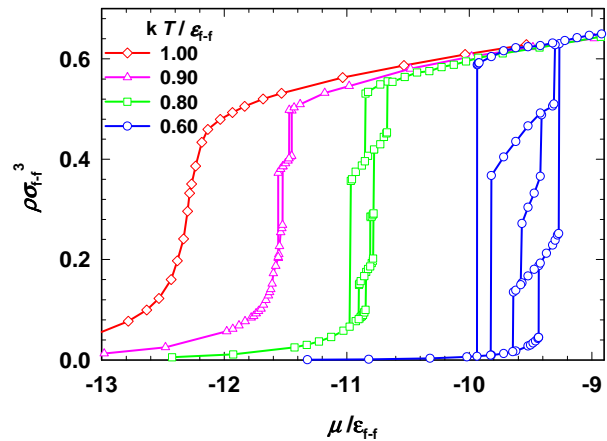, width=8cm,clip=}
\caption{Stables and metastables states obtained by GCMC simulation for various temperatures (symbols). Lines are guides to the eye indicating the path followed by the system beyond the stability limits of each metastable state. }
\label{Figisothall}
\end{figure}

Quite noticeably, the simulation points may be grouped into branches. For instance, for a reduced temperature of $0.9$, three branches are present: the gas-like branch (with a fluid layer adsorbed at the wall, the rest of the pore being filled with gas), which exists down to very low $\mu$, the liquid-like branch (pore filled by dense fluid) at high $\mu$, and a branch of intermediate density which exists only for a limited range of $\mu$. We have found that adsorption and desorption are reversible along any given branch. These branches are associated to deep local minima in the free energy landscape describing the system (the grand potential). The energetic barriers separating these minima are generally large compared to the thermal fluctuations sampled by the Monte-Carlo algorithm: the system then remains trapped in these local minima, which explains the reversibility of adsorption/desorption along the branches. However, for some particular values of $\mu$ the barriers become sufficiently small for allowing thermal fluctuations to make the system jump into an adjacent local minimum. These values define the limit of stability of the branches (in the grand canonical ensemble). The vertical lines in Fig. \ref{Figisothmain} indicate the new local minimum (branch) reached by the system. Note that the $\mu$-range of the various branches overlap, which means that for some $\mu$, the system may be stabilized in GCMC at various degrees of pore filling $\rho$. This is the main origin of hysteresis. As can be seen, the lower the temperature, the larger the $\mu$-range of existence of the branches, and the wider the hysteresis. The number of branches increases for decreasing temperature, at least down to $T ^*= 0.8$ (Fig. \ref{Figisothmain}). We shall show later that this is actually still true down to $T^* = 0.6$ if one takes into account metastable states that are hidden in the GCMC. For $T^*= 0.9$, the three branches are visited during both adsorption and desorption. They belong to the main adsorption and desorption curves. For $T^* = 0.8$, five metastable states belong to the main adsorption curve, and three to the main desorption curve. The branch of intermediate density reached during desorption (second branch, $\rho^*=\rho \sigma_{ff}^3 \sim 0.4$) looks like a continuation of the third branch visited during adsorption ($\rho^* \sim 0.4$). This point was checked by showing the reversibility of the complete branch by GCMC (by increasing and decreasing $\mu$). The total number of metastable states is then five, and their range of existence was determined in a systematic way by $\mu$-variations. The results are shown in Fig. \ref{Figisothall}. Here again, lines are guides to the eye, and the vertical lines show the new metastable state reached by the system beyond the stability limit. The case $T^* = 0.6$ deserves special attention. The procedure that consists in following simple ascending and descending $\mu$-paths does not allow to reach the two branches with reduced densities around $0.3$ and $0.45$. These states were obtained from the corresponding states at higher temperature $T^* = 0.8$ by following a particular ($\mu $,$T$) path along which their stability is preserved. This path essentially consists of slowly decreasing the temperature and increasing the chemical potential in order to keep the average number of particles roughly constant. When $T^*= 0.6$ has been reached, the $\mu$-range stability can be determined for both states (see Fig. \ref{Figisothall}). It is possible that more metastable states actually exist, but the procedures previously explained did not allow finding more than five branches in our simple system. 

\subsection{Metastable states}
\begin{figure}[ht]
\epsfig{file= 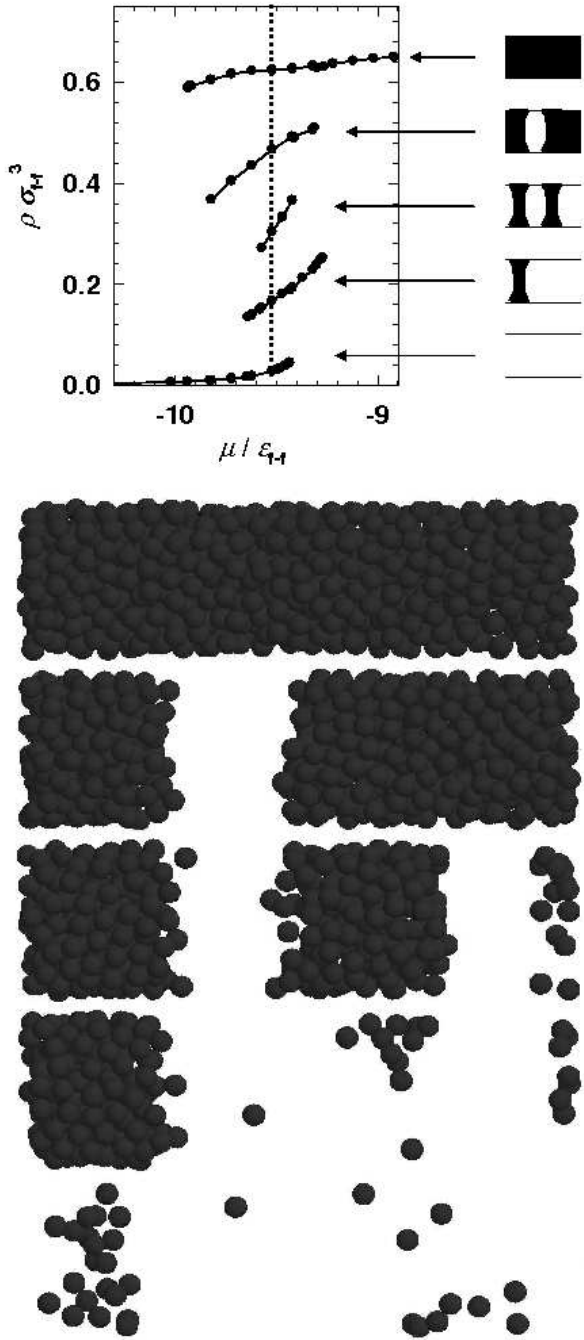, width=8cm,clip=}
\caption{Enlargement of the low-temperature metastable states found by GCMC. Solid lines indicate reversible branches. The lower panel shows at $\mu^*=-9.52$ (dotted line), for each of the five metastable states sketched in the upper panel,  one typical molecular configuration.  }
\label{Figisothenlarg}
\end{figure}

Figure \ref{Figisothenlarg} displays an enlargement of the lowest temperature ($T^*= 0.6$) GCMC results, without the vertical lines. As previously mentioned, the simulation points group into reversible branches (solid lines are guides to the eye) which correspond to local minima in the free-energy landscape. We now give a molecular-level description of these local minima. The lowest density branch corresponds to gas-like fluid filling the pore with adsorbed molecules at the wall. The highest density branch corresponds to liquidlike fluid saturating the pore. The three other states are stabilized by the chemical corrugation. Visual inspection of molecular configurations (see lower panel of Fig. \ref{Figisothenlarg}) shows that the three branches correspond respectively to: one liquid-like meniscus in the most attractive domain of the pore; two liquid-like menisci in the two attractive regions of the pore; one single large meniscus bridging the two attractive domains.

\subsection{Mixed-ensemble results}
\begin{figure}[ht]
\epsfig{file= 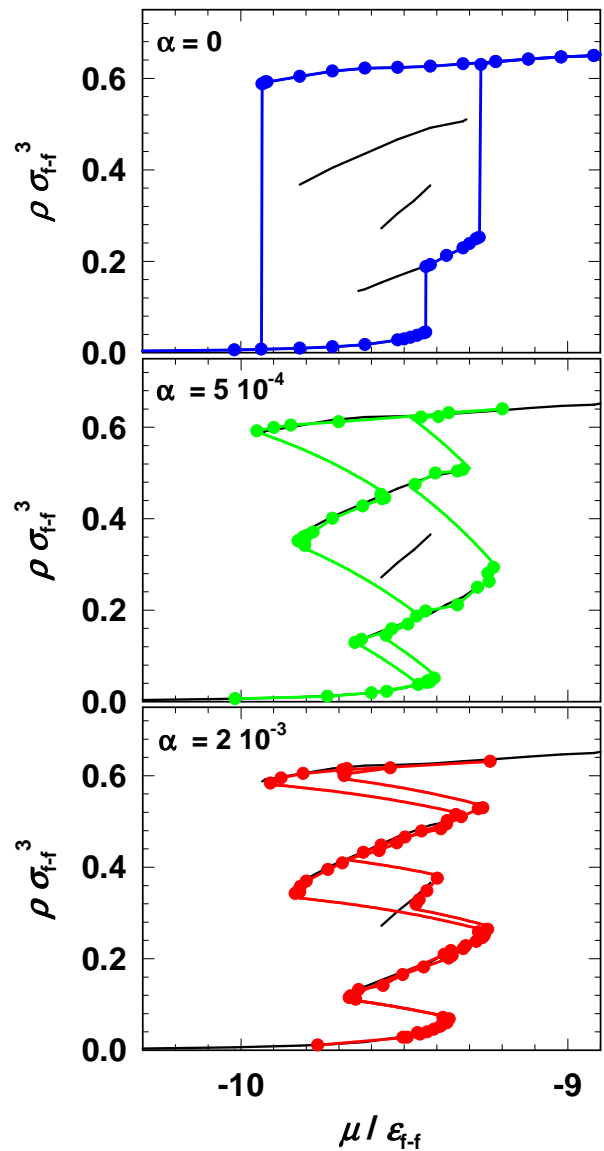, width=8cm,clip=}
\caption{Adsorption/desorption isotherms for various values of the relative size of the gas reservoir: $\alpha = 0$, $\alpha \sim 5. \times 10^{-4}$ and $ 2. \times 10^{-3}$, where $\alpha=V_P/(V_P + V_R)$. }
\label{Figisothalpha}
\end{figure}

We now focus on the adsorption/desorption results obtained for a finite-size reservoir. Two ratios $V_R /V_P = 500$ and $2000$ have been considered ($\alpha \simeq 2. \times 10^{-3}$ and $5. \times 10^{-4}$). The procedure previously used to obtain all metastable states will be applied here in its simplest form, \textit{i.e.} isothermal $\mu$-paths, because it corresponds to what is actually feasible for realistic mesoporous substrates and in real experiments (one essentially measures the main adsorption/desorption hysteresis and scanning curves). The results for the main hysteresis loops are given in Fig. \ref{Figisothalpha}, where the grand canonical case ($\alpha= 0$) has also been plotted for comparison. In all cases the symbols are the simulation data, and the color (thick) lines are guides to the eye that connect the simulation points in the same order as they were obtained by slowly increasing (adsorption) or decreasing (desorption) the total number of particles in the system. The thin (black) lines are the five branches previously obtained by extensive GCMC study (Fig. \ref{Figisothenlarg}). In the limit of infinite reservoir size ($\alpha= 0$), starting from the empty system and gradually increasing $\mu $ results in the continuous filling of the system up to $\mu^*=\mu/\epsilon_{ff}=-9.43$ where the system jumps into the second local minimum. Increasing further the chemical potential makes the system jump directly to the fifth local minimum. Upon desorption, the system remains in the saturated state for a while (hysteresis) and finally jumps (for $\mu^* = -9.94$) directly into the first local minimum (gas-like branch) without visiting any intermediate state. The vertical lines represent the constraint imposed by the reservoir ($\mu^*$= cte) but not necessarily the path actually followed by the system during the transition (transient phenomenon are not properly described by the Monte-Carlo algorithm).

In the cases of finite reservoir, it is the total number of particles in the system plus its reservoir which is increased by small steps as in a real experiment. As can be seen from Fig. \ref{Figisothalpha}, the amount of adsorbed fluid initially increases along the gas-like branch in a continous manner until it reaches its stability limit. (Note that this limit occurs at slightly larger $\mu $ for smaller reservoir: $\mu^* $= -9.43 for $\alpha =0$, $\mu^* = -9.41$ for $\alpha \simeq 5. \times 10^{-4}$ and $\mu^* = -9.36$ for $\alpha \simeq 2. \times 10^{-3}$.) At this stability limit, a small addition of extra molecules in the reservoir destabilizes the adsorbed fluid which then jumps into the second state. During the transition, particles leave the reservoir and adsorb into the pore in the most attractive region to form a meniscus (second metastable state), which results in a decrease of the chemical potential. Here again, our algorithm is not meant to describe this transition. The path actually followed by the system is not known. The (curved) line shown in the figure corresponds to the constraint of conservation of the total number of particles in the system plus its reservoir. After equilibration, the system reaches the second branch. Further increase in the total number of particles makes the system follow this branch until it reaches a new stability limit. (As previously, this limit slightly increases for a finite reservoir: $\mu^* = -9.27$ for $\alpha = 0$, $\mu ^*$ = -9.23 for $\alpha \simeq 5. \times 10^{-4}$ and $\mu^* = -9.24$ for $\alpha \simeq 2. \times 10^{-3}$.) The system then jumps onto the fifth branch for $\alpha = 0$, onto the fourth branch for $\alpha \simeq 5. \times 10^{-4}$, and onto the third branch for $\alpha \simeq 2. \times 10^{-3}$. 

\begin{figure}[ht]
\epsfig{file=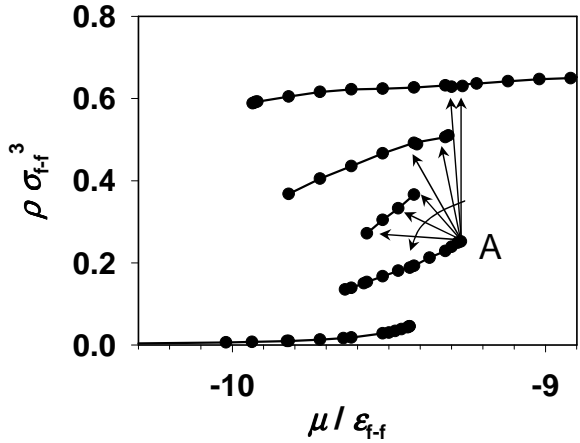, width=8cm,clip=}
\caption{Circles: GCMC results for $T^*=0.60$. Lines are guides to the eye. The arrows indicate the jump made by the representative point of the adsorbed fluid for various reservoir sizes, after destabilization of the initial point A by a small increase in the total number of particles in the system. The curved arrow indicates increasing values of the relative reservoir size: $\alpha$ = 0. (GCMC), $2.2 \times 10^{-4}, 2.5 \times 10^{-4}, 5.0 \times 10^{-4}, 1.0 \times 10^{-3}, 2.0 \times 10^{-3}, 2.4 \times 10^{-2}$.}
\label{Figjumpalpha}
\end{figure}

\begin{figure}[ht]
\epsfig{file= 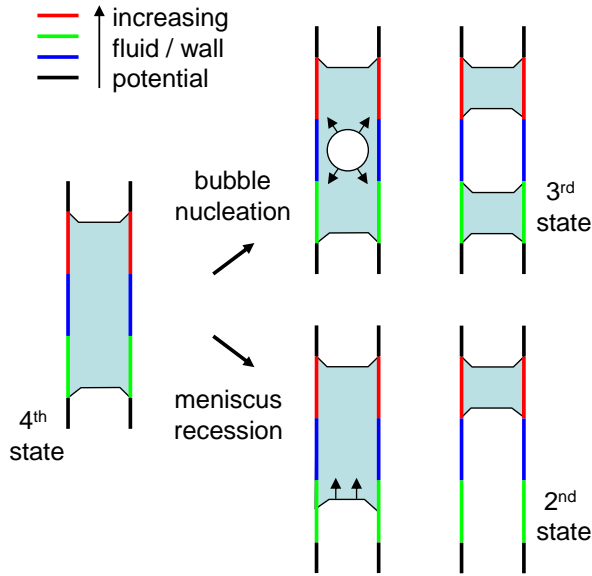, width=8cm,clip=}
\caption{Possible mechanisms for the evolution of the fluid configuration inside the pore starting from a given metastable branch (see text).}
\label{Figmechanism}
\end{figure}

The above results illustrate one of the main effects of the reservoir: the system does not necessarily visit the same metastable states as one changes the size of its reservoir. From Fig. \ref{Figisothalpha}, this effect may be tentatively interpreted as follows: the reservoir imposes a constraint on the relation between the chemical potential and the amount adsorbed, materialized by the vertical or inclined curves. After a branch stability limit has been reached, the system approximately follows this line of constraint during the transition, and finally meets a new branch at a location that depends on the ratio between the reservoir and the system size. To illustrate this point, starting from the highest $\mu$ GCMC point of the second branch (limit of stability), we have determined the new metastable state reached by the system after a small increase in the total number of molecules for various reservoir sizes. The results are given in Fig. \ref{Figjumpalpha}. This picture is however somewhat oversimplified and does not apply in all situations since it does not take into account the complexity of the underlying free-energy (grand-potential) landscape that determines which path is actually followed by the system (being in configurational space, the free energy landscape is an object of very high dimension). At a molecular level, this means that complex nucleation and fluid adsorption processes determine the path actually followed by the system. One of the consequences is that the system does not necessarily stop on the first branch crossed by the constraint line. This is illustrated in our simple system for $\alpha \simeq 2. \times 10^{-3}$ during desorption from the fourth branch: as can be seen, the constraint line crosses the third metastable branch, but the system avoids this state and actually reaches the second branch. In the grand-potential landscape picture, this means that, starting from the fourth local minimum, and following the steepest slope, the system has not crossed the basin of attraction associated to the third minimum and has followed its way until being finally trapped by the basin associated to the second local minimum. In the molecular-level description, the fourth branch corresponds to a large meniscus bridging the two attractive regions. Upon desorption, the system would have to nucleate a bubble in between the attractive regions in order to reach the third branch corresponding to two menisci in the attractive regions (see Fig. \ref{Figmechanism}, upper path). This nucleation barrier is probably too high, and it is more favorable for the system to desorb by recession of the meniscus in the less attractive regions until it is left with one single meniscus in the most attractive region corresponding to the second state (see Fig. \ref{Figmechanism}, lower path).

\begin{figure}[ht]
\epsfig{file= 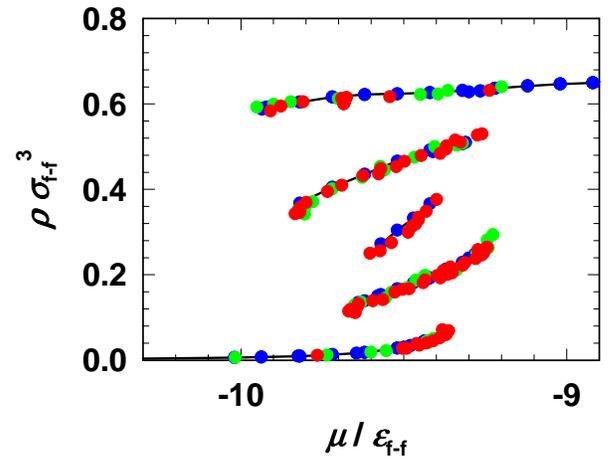, width=8cm,clip=}
\caption{Metastable states of the adsorbed fluid obtained for all the different sizes of the reservoir studied here (shown with different colors; colors on line).}
\label{Figallmetasalpha}
\end{figure}

As for the grand-canonical case, it is possible to perform a systematic search for all possible metastable states in the system in contact with a finite reservoir. The five branches already described are found in all cases. Figure \ref{Figallmetasalpha} shows the superimposition of all simulation points (stable over long runs) obtained for the various ratios between the reservoir and system sizes already presented. As can be seen, the points group onto branches corresponding to the five metastable states found in the system. It is however reminded that the limits of stability of the branches slightly vary with the reservoir size. This may be related to the fact that transitions proceed via energetic barriers and that the amplitude of the fluctuations allowed by the reservoir depends on its size. To summarize the results, reducing the size of the reservoir allows one to explore metastable states upon adsorption/desorption that are not accessible in the  grand-canonical situation. It is also found that the number of metastable states visited by the fluid increases with decreasing reservoir size.

\section{Results for the coarse-grained model}

\subsection{Grand-canonical isotherms}
\begin{figure}[ht]
\epsfig{file=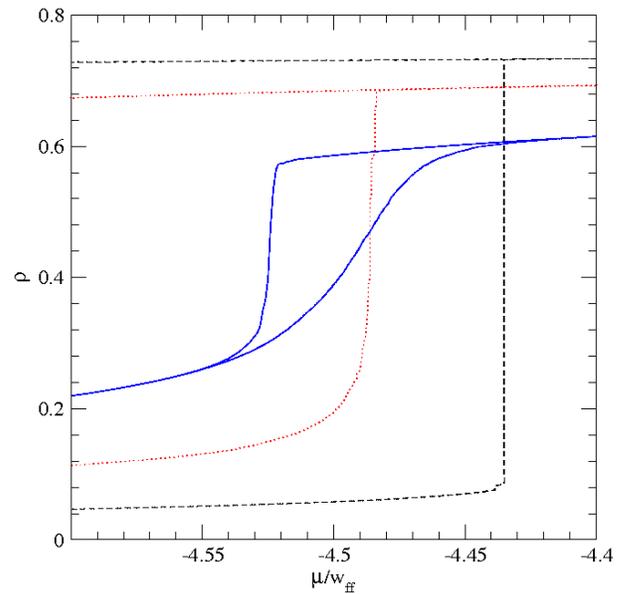, width=8cm,clip=}
\caption{ Hysteresis loops around the ``capillary condensation'' in the grand canonical ensemble for a $L=100$ sample as a function of the temperature: $T^*=1.40$ (full blue line), $T^*=1.10$ (dotted red line), $T^*=0.80$ (dashed black line) (colors on line). For the last two temperatures, only part of the desorption branch is shown.}
\label{FigEK1}
\end{figure}

\begin{figure}[ht]
\epsfig{file=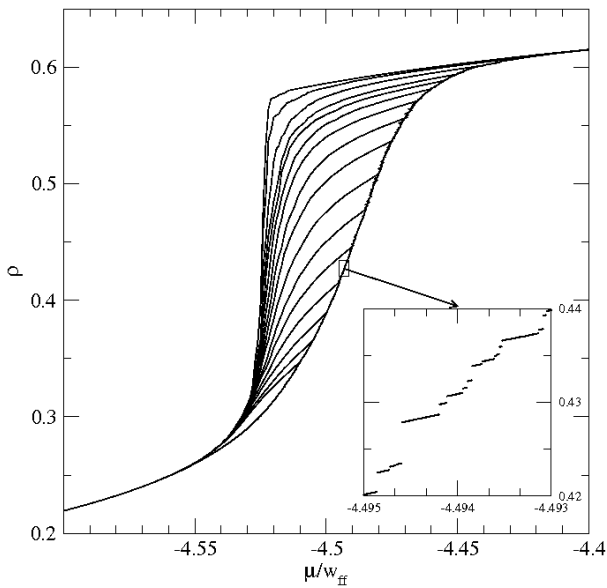, width=8cm,clip=}
\caption{ Grand canonical descending scanning curves in a $L=50$ sample at $T^*=1.40$. A close-up of the adsorption isotherm is shown in the inset.  }
\label{FigEK2}
\end{figure}

Typical results for adsorption isotherms obtained at various temperatures in the grand-canonical ensemble are shown in Fig. \ref{FigEK1}. When the temperature decreases, the isotherm shape changes from smooth to steep. As discussed in previous papers,\cite{RKT2003} this corresponds to a true out-of-equilibrium phase transition, the so-called ``avalanche transition'',\cite{SDKKRS1993} with the sudden appearance of a macroscopic, connected liquid domain in the whole porous sample. A previous extensive scaling study\cite{RKT2003} showed that it exists a critical value of the temperature for which the isotherm changes from continuous to discontinuous in the thermodynamic limit. At higher temperature, isotherms look gradual but at reduced temperatures $T^*=1.40$, they consist of little steps of varying sizes (see insert in Fig. \ref{FigEK2}).

The above results are related to the characteristics of the grand-potential landscape. At low temperatures, this multidimensional landscape (recall that it is a function of the local fluid densities, with here $N_{SP}\sim 10^4-10^6$) is characterized by a large number of local minima, the metastable states in which the system may be trapped. Since thermally activated processes are neglected in the LMFT (an approximation that, as already discussed, finds its justification in the experimental reproducibility of the adsorption isotherms on the time scale of most experiments), the evolution of the system is only due to a variation of the external driving (here the chemical potential). As $\mu $ varies, the system either follows the minimum in which it is trapped as this minimum deforms gradually (the flat portions in the insert of Fig. \ref{FigEK2}), or it falls instantaneously into another minimum when the former reaches its stability limit. This later move is a discontinuous and irreversible process, an avalanche, which is at the origin of the history-dependent behavior of the system, \textit{e.g.} the hysteresis. The avalanche corresponds to some collective condensation event inside the porous sample that manifests itself by a jump in the adsorption isotherm. The size of the avalanche may be macroscopic as for $T^*=0.8$, and the adsorption isotherm is discontinuous in the thermodynamic limit; or it may be microscopic as for $T^*=1.40$, and the adsorption isotherm remains smooth in the thermodynamic limit. (More calculations should be necessary to conclude on what happens at $T^*=1.10$.) The hysteresis loop encloses all the metastable states of the system: this is illustrated in Fig. \ref{FigEK2} by the scanning curves obtained by performing incomplete filling of the matrix and then decreasing the chemical potential to drain the adsorbed fluid. Of course, the number of metastable states is very large, probably exponential in the system size, and only few of them are revealed with this simple procedure.

\subsection{Canonical and mixed-ensemble isotherms}
\begin{figure}[ht]
\epsfig{file=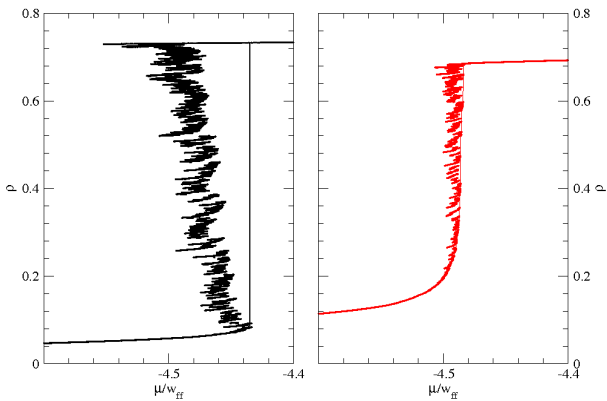, width=8cm,clip=}
\caption{ Canonical (symbols) and grand-canonical (lines) adsorption isotherms in a $L=100$ sample at $T^*=1.10$ (right) and $T^*=0.80$ (left).   }
\label{FigEK5}
\end{figure}
\begin{figure}[ht]
\epsfig{file=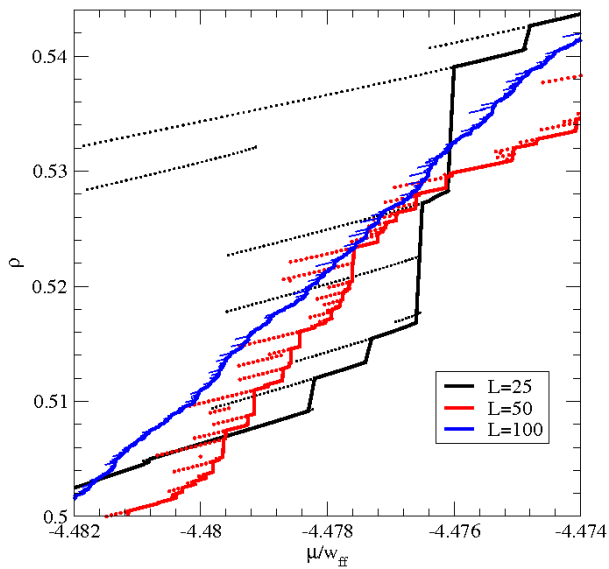, width=8cm,clip=}
\caption{A close-up of the canonical (symbols) and grand-canonical (lines) adsorption isotherms at $T^*=1.40$ for different linear sizes of the system (colors on line).  }
\label{FigEK4}
\end{figure}

What happens when the porous sample is coupled to a finite reservoir? First, we begin with a vanishingly small reservoir, which is the canonical situation where one controls the number of adsorbed particles. The main results are summarized in Fig. \ref{FigEK5} where we compare the adsorption isotherms obtained with the ``$\mu$-driven'' (grand-canonical) and ``$\rho$-driven'' (canonical) procedures. The behavior is quite different in the low and the high temperature regimes, respectively characterized by the absence and the presence of a macroscopic $\mu$-driven avalanche. At the highest temperature studied ($T^*=1.80$, not shown in the figure), there exists only one stable state, the equilibrium one, and controlling either the adsorbed density or the chemical potential yields the same result. This is also true for all temperatures at very low adsorbed densities (not shown in the figure). At lower temperatures, avalanches appear in the grand-canonical  isotherm. Collective localized events, which we define as ``avalanches'' irrespective of the control variable, also show up in the canonical isotherm in the form of small jumps in the chemical potential toward a lower value: with the disappearance of the initial minimum, the system has to find another metastable state with the required adsorbed density; as this state cannot be found at a higher value of the chemical potential since the grand-canonical adsorption/desorption isotherms have been shown to represent the extremal curves that encompass all the metastable states of the system,\cite{KMRT2002} the chemical potential must decrease. (Note that in a grand-canonical setting ``avalanches'' appear as jumps in the adsorbed density $\rho$ at constant chemical potential $\mu$ whereas in a canonical one, they appear as jumps in $\mu$ at constant $\rho$.) When a metastable state is found and fluid is further added, the system smoothly follows this state (see the quasi-linear portions in Fig. \ref{FigEK4}) until it reaches the corresponding stability limit. Then, there is a new jump in the chemical potential to a smaller value, and the evolution proceeds in this way until the porous sample is completely filled with liquid. The contrast between $\mu$-driven and $\rho$-driven isotherms that is illustrated in Fig. \ref{FigEK5} is very reminiscent of what was found in ferromagnetic systems when comparing magnetization-driven and magnetic-field-driven protocols.\cite{IRV2006} As in the latter case, the canonical ($\rho$-driven) isotherms are closely related to the distribution of metastable states inside the (grand-canonical) hysteresis loops.

We now discuss in more detail the canonical ($\rho$-driven) isotherms. At $T^*=1.40$ with a sample of linear size $L=100$, the jumps in $\mu(\rho)$  are very small and one needs to zoom in on the isotherm to see small differences between the canonical and grand-canonical protocols, as illustrated in Fig. \ref{FigEK4}. This smallness is related to the presence of many metastable states in the close vicinity of the grand-canonical isotherm. Since both the grand-canonical and the canonical avalanches remain of microscopic size, we expect that the small jumps in either $\rho$ or $\mu$ become infinitesimally small in the thermodynamic limit. This is indeed what is observed in Fig. \ref{FigEK4} where isotherms obtained for different system sizes are compared. Therefore, we predict that the canonical and grand-canonical curves should become identical in the thermodynamic limit.

\begin{figure}[hb]
\epsfig{file=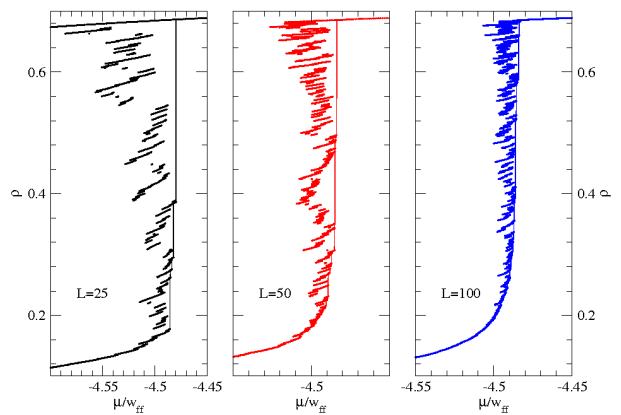, width=8cm,clip=}
\caption{Canonical (symbols) and grand-canonical (lines) adsorption isotherms at $T^*=1.10$ for samples of different sizes: $L=25$ (left), $L=50$ (middle), $L=100$ (right).}
\label{FigEK6}
\end{figure}

A similar behavior is observed at $T^*=1.10$, as illustrated in Fig. \ref{FigEK6}. However, the jumps in $\rho$ are larger than at $T^*=1.40$ and are clearly associated with a much more fluctuating chemical potential $\mu(\rho)$: each jump in $\rho$ generates a re-entrant behavior in $\mu(\rho)$. It is difficult to conclude on how the isotherm will evolve in the thermodynamic limit: as the size of the system increases (see Fig. \ref{FigEK6}), there are more and more intermediate points in the steepest part of the grand-canonical isotherm and, at the same time, the canonical isotherm is less and less fluctuating. The temperature $T^*=1.10$ is probably barely above the critical temperature of the (grand-canonical) avalanche transition and we expect that both the canonical and the grand-canonical isotherms would become continuous and would coincide in the thermodynamic limit.

\begin{figure}[ht]
\epsfig{file=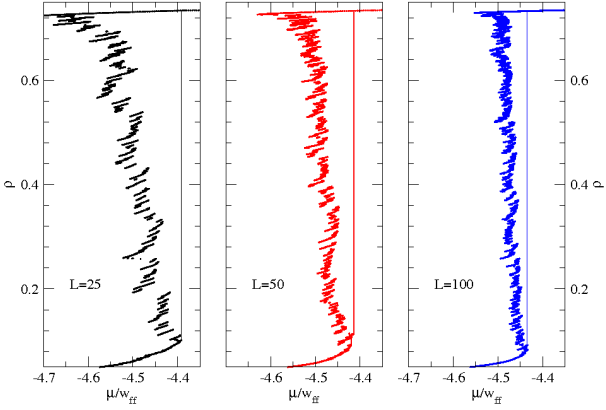, width=8cm,clip=}
\caption{Canonical (symbols) and grand-canonical (lines) adsorption isotherms at $T^*=0.80$ for samples of different sizes: $L=25$ (left), $L=50$ (middle), $L=100$ (right).}
\label{FigEK7}
\end{figure}

On the other hand, $T^*=0.80$ is undoubtedly below the critical temperature of the avalanche transition: all samples studied, whatever their size, display a large jump in their grand-canonical isotherms (see Fig. \ref{FigEK7}).  (In a previous work on a random matrix,\cite{RKT2003} we concluded through an extensive finite-size study that a macroscopic jump exists in the thermodynamic limit at $T^*=0.8$ for $y=0.9$; since the value $y=0.8$ used in this paper corresponds to a weaker disorder than $y=0.9$, this is necessarily true here as well.) As shown in Fig. \ref{FigEK7}, the canonical curves show a pronounced reentrant behavior, with a large difference with the grand-canonical ones: a whole region void of metastable states then appears. There are always finite-size effects, which affect both the canonical and the grand-canonical isotherms. In the latter case, the rare events that trigger the macroscopic avalanches are very sensitive on details about the structure of the matrix, and the corresponding chemical potentials vary strongly with the system size at small $L$. We have nonetheless checked by performing grand-canonical isotherms on very large systems of linear size $L=200$ that the position of the jump in the thermodynamic limit is very close to the jump found in the isotherm for $L=100$ that is displayed in Fig. \ref{FigEK7}. On the other hand, even if the chemical potentials of the  canonical isotherms fluctuate less and less as the size of the system grows (especially at the end of the adsorption process, when the porous sample is nearly filled with liquid), the overall location of the isotherm in the $\rho$-$\mu$ plane does not change significantly with $L$.

\begin{figure}[ht]
\epsfig{file=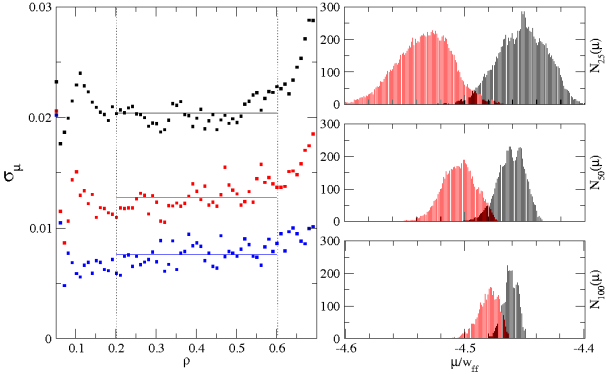, width=8cm,clip=}
\caption{Left: Standard deviations $\sigma_{\mu}$ of the chemical potentials as a function of the adsorbed density $\rho$ at $T*=0.80$ for different system sizes: $L=25$ (black), $L=50$ (red), $L=100$ (blue). Lines show averaged values. Right: histograms of $\mu(\rho)$ for different values of $\rho$ ($\rho=0.2$ in red/left, $\rho=0.6$ in black/right ) and different system sizes ($L=25$: top, $L=50$: middle, $L=100$: bottom). }
\label{FigEK9}
\end{figure}

\begin{figure}[ht]
\epsfig{file=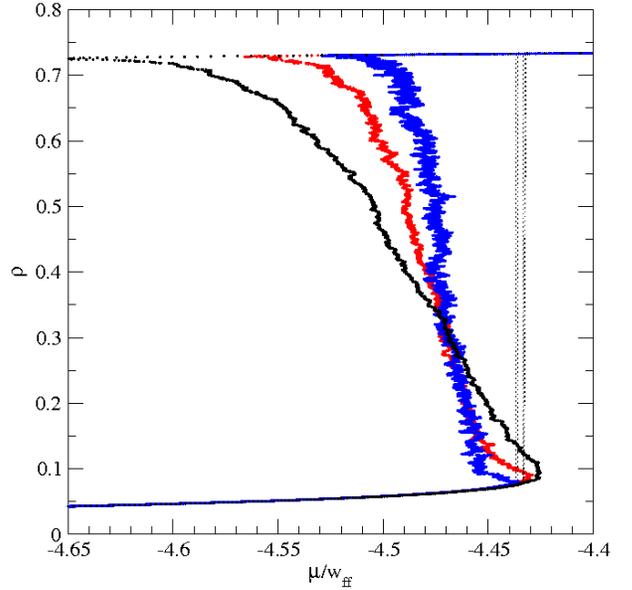, width=8cm,clip=}
\caption{Average canonical adsorption isotherms $\overline{\mu(\rho)}$ at $T^*=0.80$ for different system sizes; from left to right in the upper part of the plot (colors on line): $L=25$ (black), $L=50$ (red), $L=100$ (blue). The corresponding numbers of realizations are 256, 64 and 8, respectively. Typical grand-canonical adsorption isotherms for $L=200$ samples are also shown (dashed curves). }
\label{FigEK8}
\end{figure}

To discuss the fluctuations of the chemical potential with the density in the canonical protocol, we have performed finite-size studies of the first moments of their distribution, after averaging over disorder (\textit{i.e.} porous sample) realizations. More precisely, we compute the mean $\overline{\mu(\rho)}$ (hereafter an overbar denotes the average over disorder) and the variance $\Delta_{\mu} (\rho)=  \overline{\mu(\rho)^2}- \overline{\mu(\rho)}^2$ for three sizes of the system. As explained in Ref. \cite{IRV2006}, the standard  argument concerning self-averaging quantities, whose value in a macroscopic sample is equal to the average over all disorder realizations, cannot be applied: as a consequence, one cannot be sure that $\mu(\rho)$ is a self-averaging quantity. To delve more into its behavior, it is interesting to investigate the dependence on system size of the standard deviation $\sigma_{\mu}(\rho)=\sqrt{\Delta_{\mu}(\rho)}$. The result is shown in Fig. \ref{FigEK9},  where $\sigma_{\mu}(\rho)$ is plotted as a function of the adsorbed density (computed in bins of width 0.01). Except at low and high density, the standard deviation remains nearly constant with $\rho$  (typically, between $\rho=0.2$ and $\rho=0.6$). It decreases with $L$ but quite slowly: we find that in the region between $\rho=0.2$ and $\rho=0.6$, $\Delta_{\mu}(L)=\sigma_{\mu}(L)^2 \sim L^{-\gamma}$ with a finite-size scaling exponent $\gamma$ around 1.4, \textit{i.e.} significantly smaller than 3. This indicates that $\mu(\rho)$  is self-averaging, but only weakly so. As in Ref. \cite{IRV2006}, we have checked that the corresponding histograms are roughly Gaussian (see Fig. \ref{FigEK9}). Although the situation with respect to self-averaging remains somewhat unclear as far as the whole isotherm is concerned,\cite{IRV2006} it is nonetheless instructive to study the mean $\overline{\mu(\rho)}$ as a function of system size. This is displayed in Fig. \ref{FigEK8}, where one can see that the $\overline{\mu(\rho)}$-isotherms vary significantly with the size of the system, becoming steeper as $L$ increases. We cannot conclude whether or not the isotherm becomes vertical in the thermodynamic limit, but the data strongly suggest that it will be distinct in this limit from the grand-canonical isotherm: the  $\overline{\mu(\rho)}$-isotherms for all studied system sizes intersect near the point $\mu^*=-4.47$ and $\rho=0.3$, at a significantly lower chemical potential than those found for the grand-canonical jumps ($\mu^* \sim -4.43 -4.44$) for $L=200$ (see Fig. \ref{FigEK8}).

\begin{figure}[ht]
\epsfig{file=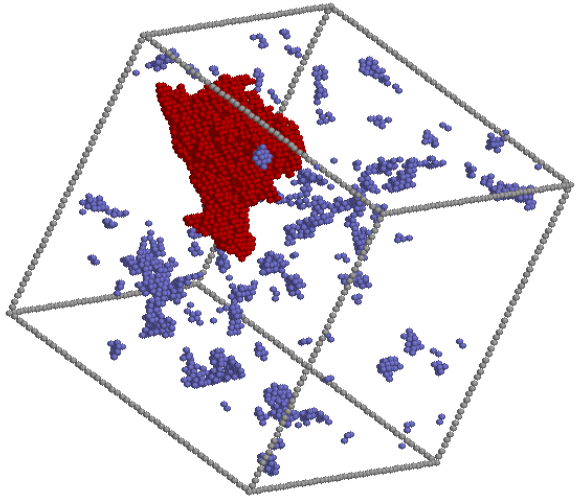, width=8cm,clip=}
\caption{Condensation and evaporation events between $\rho=0.3962$ and $\rho=0.3963$ in the $\rho$-driven protocol.}
\label{FigEK11}
\end{figure}

We now comment on the two types of behavior found along the $\rho$-driven adsorption isotherm. For increasing adsorbed density $\rho$, the chemical potential may either increase continuously or decrease by discontinuous jumps. In the former case, there are only slight swellings of the liquid domain, with the liquid-gas interfaces retaining the same shapes locally and all the $\rho_i$'s increasing. In the latter case, we find that each jump corresponds to a single condensation event,  a $\rho$-driven ``avalanche'', in which all sites of a compact region become liquid. This is illustrated in Fig. \ref{FigEK11}. Considering the two consecutive configurations with average densities $\rho$ and $\rho+\Delta \rho$ (before and after the jump),  sites are considered as turning liquid (respectively, gas)  when the variation of the local fluid density  is larger than $0.3$ (respectively, lower than $-0.3$) and are shown in red (respectively, grey) in the figure. The size of the condensed region varies for each jump along the isotherm, but the number of particles involved in the local condensation can go much beyond the controlled increment $V_P \Delta \rho$: in Fig. \ref{FigEK11}, the  condensation event (in red) corresponds to a local  increase in the number of particles that is $110$ times larger than the initial increment. Therefore, others regions of the porous sample must be drained: as shown in the figure, the most important evaporation events (in grey) remain smaller and less compact than the condensation event and appear at random throughout the material. (Whereas the definition of  ``turning liquid'' is well characterized since the histogram of positive variations of the local density is always peaked at a value close to $1$, this is much less so for its counterpart ``turning gas'':  the histogram of negative variations is continuously decreasing between $0$ and $-1$, so that the decrease of the average density that balances the condensation event mostly comes from small but widespread local variations.) This suggests the following interpretation: adding a small amount of fluid in the sample may trigger a condensation ``avalanche'' whose size is larger than the added fluid amount; a decrease of the chemical potential allows a slight draining of the rest of the sample and provides the required amount of fluid to ``feed'' the avalanche.

\begin{figure}[ht]
\epsfig{file=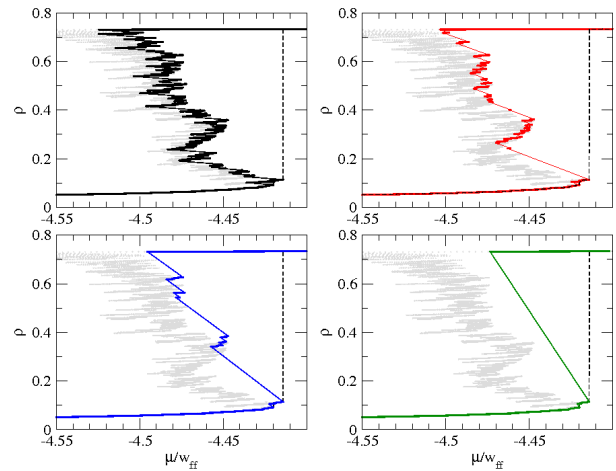, width=8cm,clip=}
\caption{Adsorption isotherms at $T^*=0.80$ for different relative sizes of the reservoir for a $L=50$ sample. In the four panels, the canonical isotherm is shown in light grey and the grand-canonical isotherm as a dashed curve. Top left: $\alpha =10^{-2}$, top right: $\alpha =2. \times 10^{-3}$, bottom left: $\alpha =10^{-3}$, bottom right:  $\alpha =5. \times 10^{-4}$, where $\alpha=V_P/(V_P + V_R)$. Note that both the slope and the length of the inclined segments that represent jumps in the $\mu$-$\rho$  plane (the lines are therefore a guide to the eye) increase as $\alpha$ decreases: the slope is directly related to $\alpha$ (compare with Fig. \ref{Figjumpalpha}). }
\label{FigEK10}
\end{figure}

Finally, Fig. \ref{FigEK10} presents our results in the mixed-ensemble for different sizes of the reservoir. As the size increases, the $\mu(\rho)$-isotherms exhibit flat segments with decreasing length and jumps in the  $\mu $-$\rho$  plane that are more and more pronounced. This behavior is similar, albeit in a more complex systems and in the presence of a much larger number of metastable states, to that found in the atomistic model and illustrated in Fig. \ref{Figisothalpha}. When the volume of the reservoir becomes $2000$ times greater than the volume of the porous material, the isotherm displays a unique jump between the grand-canonical low-density branch and the grand-canonical high-density branch, but it still comes with a decrease of the chemical potential (lower right panel of Fig. \ref{FigEK10}). For larger reservoir sizes, the jump gets closer and closer to the grand-canonical isotherm. Convergence however is rather slow. On the other hand, for the smallest reservoir sizes (a reservoir volume $500$ times the volume of the porous material or less: see the two upper panels of Fig. \ref{FigEK10}), the isotherms tend to superimpose on the canonical isotherm. It can be however noticed that the explored metastable states are not necessarily identical. As explained in section III, one indeed has to be very careful with the representation of states in the $\mu $-$\rho$  plane which is only a projection from the highly multi-dimensional configurational space: it is possible to cross branches as the mixed-ensemble constraint allows the system to explore other areas of the underlying free-energy landscape.

\section{Conclusion}

In this paper, we have presented a study of the out-of-equilibrium, hysteretic response of a fluid adsorbed in inhomogeneous porous materials when one couples the porous sample with a finite-size reservoir and controls the total number of particles. Varying the relative size of the sample and the reservoir allows one to interpolate between a canonical situation with a controled  adsorbed density and a grand-canonical situation with a controled chemical potential. We have considered both an atomistic model of a fluid in simple, yet structured pore and a coarse-grained model for adsorption in a disordered mesoporous material. The adsorption isotherms have been computed by molecular Monte Carlo simulation in the former case and by a density functional approach in a local mean-field approximation in the latter. The Monte Carlo simulations give us a clear picture of what occurs at the molecular scale in a small and weakly disordered system whereas the density functional approach provides insights at the mesoscopic scale for a fully disordered system. In both cases, we have found \textit{metastable states} that appear as branches of finite extent in the $\mu-\rho$ plane. The metastable states correspond to inhomogeneous configurations of the fluid and the number of branches increases rapidly with the complexity of the material. It is worth stressing two points: first, these states are not unstable, \textit{i.e.} not purely stabilized by the constraint on the total number of particles\cite{E1998}; second, metastability is induced by the intrinsic inhomogeneity of the solid. Accordingly, the picture of metastability in such systems is quite different than that encountered in mean-field-like descriptions of homogeneous (bulk) systems undergoing first-order transitions.

We have shown that the way the system evolves between these metastable states may depend on the protocol, controlled here by the relative size of the reservoir. In particular, our results suggest that a discontinuity in the grand-canonical adsorption isotherm (an out-of-equilibrium ``avalanche transition'') is associated with the absence of metastable states in a whole region of the $\mu-\rho$ plane and that the corresponding canonical adsorption isotherm (and, more generally, isotherms performed with a small enough reservoir size) differs from the grand-canonical one and displays a reentrance, even in the thermodynamic limit.

What is the relevance of our results for experimental set-ups ? For illustration, we consider two examples. In order to determine the adsorption isotherm with the volumetric method using for instance nitrogen, known amounts of fluid are admitted stepwise in the sample cell. The amount of adsorbed fluid is the difference between the admitted gas and the amount of gas that fills the ``dead volume'', \textit{i.e.} the free space in the sample cell including connections: this is equivalent to the procedure discussed in this paper, with the dead volume playing the role of a finite-size reservoir for the sample. The amount of gas in this reservoir is calculated from the fluid equation of state and from measurements of the pressure, the temperature and the dead volume. In an experiment with a sample size of 20 mm$^3$, a cell of a few cm$^3$ could be used. In such a situation, the ratio of volumes (porous medium versus total) is around $10^{-2}$ and the ratio of adsorbed amount on the total amount around $10^{-1}$. On the other hand, to measure Helium adsorption in aerogels, E. Wolf et al \cite{BPW2007} use a different experimental set-up that was first proposed by Herman et al\cite{HDB2005}, in which the adsorbed amount is controlled through the temperature of a helium gas reservoir connected to the experimental cell.  The total amount of Helium is fixed, but varying the temperature of the reservoir transfers atoms from the reservoir to the cell, or conversely. In the recent experiments,\cite{BLCGDPW2008} with temperatures ranging from 4K to 5K, the adsorbed amount at saturation corresponds to around 20\% of the total amount.

In both above examples, the ratio between the adsorbed amount and the total amount of fluid is fully in the range of the parameters of our study. However, in most experiments in disordered porous materials the adsorption isotherms display hysteresis but are smooth and continuous: in consequence, as predicted by the present study, no effects of the reservoir size are expected for macroscopic systems. We have shown here that for such effects to be observable, the temperature should be low enough for the grand-canonical adsorption isotherm to exhibit a true discontinuity and not only a very steep variation. It is unlikely that this true discontinuity could be observed in common disordered porous materials such as Vycor or xerogels: the disorder and the confinement are too strong and prevent the appearance of the avalanche transition at temperatures higher than the triple point. The case of helium adsorption in aerogels is however more promising since at low temperature (below 4K), jumps have been predicted\cite{DKRT2004,DKRT2005}and may have already been observed in some experiment\cite{TYC1999}. It would therefore be interesting in this case to perform controlled experiments for different relative sizes of the reservoir to see if the predicted reentrance of the isotherm as one moves away from the grand-canonical situation is indeed encountered, as observed in hysteretic martensitic transformations.\cite{BRIMPV2007}

\acknowledgments
We thank M.-L. Rosinberg for many fruitful discussions.

\appendix*

\section{DFT Algorithms}
As a starting point, it is convenient to rewrite  the condition of minimization of the grand potential, $\frac{\partial \Omega_T}{\partial \rho_i}=0$, as
\be
\label{EqA1} 
\exp (-\beta \lambda ) \rho_i=  (\eta_i-\rho_i) \exp[\beta (w_{ff}\sum_{j/i} \rho_j +w_{sf}\sum_{j/i} (1-\eta_j))] 
\ee
and sum over $i$ so as to express $\lambda $ as a function of the densities 
\begin{align}
\label{EqA2} 
&\exp (-\beta \lambda ) = \nonumber \\
  &\frac{\sum_i \{ (\eta_i-\rho_i) \exp[\beta (w_{ff}\sum_{j/i} \rho_j +w_{sf}\sum_{j/i} (1-\eta_j))] \} }{\sum_i  \rho_i}.
\end{align}
Our algorithm is then the following: changing the total density $\rho_T $ by a small step $\Delta \rho$, \textit{i.e.} $\rho_T^{new}=\rho_T^{old}+\Delta \rho$,  one first supposes that the supplementary amount is adsorbed in the porous material: $\rho^{new}=\rho^{old}+\Delta \rho$. One then computes the new Lagrange parameter from
\begin{align}
\label{EqA3} 
&\exp (-\beta \lambda^{new} ) = \nonumber \\
  &\frac{\sum_i \{ (\eta_i-\rho_i^{old}) \exp[\beta (w_{ff}\sum_{j/i} \rho_j^{old} +w_{sf}\sum_{j/i} (1-\eta_j))] \} }{V_P \rho^{new}}
\end{align}
(with the old local densities) and the new local densities from
\begin{align}
\label{EqA4} 
&\rho_i^{new}= \nonumber \\
&\frac{\eta_i}{1+\exp [ -\beta (\lambda^{new} +w_{ff}\sum_{j/i} \rho_j^{old}+w_{sf}\sum_{j/i} (1-\eta_j))]}.
\end{align}
The new reservoir density  $\rho_R^{new}$ is obtained by inverting 
\be
\label{EqA5} 
\lambda^{new}=k_BT \ln [\frac{\rho_R^{new}}{1-\rho_R^{new}}]-w_{ff}\rho_R^{new}.
\ee
One then checks if the constraint is satisfied, \textit{i.e.} if $\rho_T=\alpha \frac{\sum_i \rho_i^{new}}{V_P}+(1-\alpha)\rho_R^{new}$ within a given precision and, if not, one iterates using 
$\Delta \rho=\frac{1}{\alpha}[\rho_T - \alpha \frac{\sum_i \rho_i^{new}}{V_P}-(1-\alpha)\rho_R^{new}]$ until convergence is reached.

In practice, to improve the convergence of the iteration procedure, we use a mixing scheme, retaining a part of the previous iteration for the subsequent iteration.  It appears that the configurations visited do not depend on the mixing parameter when the convergence criterions are strong enough. 

Note that other equivalent algorithms can be devised as well and, interestingly, the output of the calculation appears to be quite robust to the choice of the algorithm. For instance, changing Eq. \ref{EqA2} by the equivalent equation,
\be
\label{EqA6} 
\exp (-\beta \lambda ) =  \frac{\sum_i  (\eta_i-\rho_i)}{\sum_i \{ \rho_i \exp[\beta (w_{ff}\sum_{j/i} \rho_j +w_{sf}\sum_{j/i} (1-\eta_j))] \} },
\ee
yields the same trajectory for the converged states even if the intermediate stages and the speed of convergence greatly differ. One can also add the increments $\Delta \rho$  in the reservoir as in the molecular simulation:  $\rho_R^{new}=\rho_R^{old}+\Delta \rho$; then, one computes the new Lagrange multiplier with Eq. \ref{EqA5} and the new local densities with Eq. \ref{EqA4}, checks the constraint, and iterates using $\Delta \rho=\frac{1}{1-\alpha}[\rho_T- \alpha \frac{\sum_i \rho_i^{new}}{V_P}-(1-\alpha)\rho_R^{new}]$.  This does not change the isotherm when the evolution is adiabatic. This algorithm seems simpler and possibly closer to the experimental protocol. However, it becomes problematic in the limit of the canonical ensemble ($\alpha $=1). Therefore, we have preferred the first algorithm described above.

Our calculations were performed with samples of linear sizes varying from $L=25$ to $ L=100$ and convergence was assumed when for the $n^{\rm th}$ iteration, $\max\limits_{\{i\}} \left|\rho_i^{(n-1)}-\rho_i^{(n)}\right| < \epsilon$ and $\left|\rho^{(n-1)}-\frac{\sum_i \rho_i^{(n)}}{V_P}\right| < \epsilon$ with $\epsilon =10^{-6}$ for $L=25$ and $\epsilon =10^{-8}$ for $L=50$ and $L=100$. In addition, the steps $\Delta \rho$ in $\rho $ were taken as small as $10^{-5}$ so that most of the avalanches could be resolved (see Ref. \cite{DKRT2006} for more details about identification of avalanches). Enforcing further the convergence criterions ({\it e.g.} with $\epsilon =10^{-9}$ for $L=50$) does not change the path followed by the system: adiabatic regime has been reached. However, for the smallest system studied, with linear size $L=25$, diminishing too much $\Delta \rho$ could prevent convergence as the system is not able to find a metastable state with the required density.


\begin{thebibliography}{10}
\bibitem{fn1} In general, both the system, \textit{i.e.} the sample with the porous medium, and the reservoir are taken in the thermodynamic limit; yet the reservoir can be of negligible size compared to the system.
\end{thebibliography}
\end{document}